\documentclass[useAMS,usenatbib,usegraphicx]{mn2e} 
\usepackage{longtable} 
\usepackage{color} 
\newcommand{\ion}[2]{#1\,{\mdseries\textsc{#2}}} 
\definecolor{red}{rgb}{1.00,0.00,0.00}

\bibpunct{(}{)}{;}{a}{}{,}

\title[Optical Observations of SN Ic-BL PTF10qts]{Optical Follow-Up Observations of PTF10qts, a Luminous Broad-Lined Type Ic Supernova Found by the Palomar Transient Factory}

\author[E.S. Walker et al.] 
{
\parbox{\textwidth} 
{E.S.~Walker$^{1,2}$ \thanks{E-mail address emma.walker\@ yale.edu}, P.A.~Mazzali$^{3,4,5}$, E.~Pian$^{1,6,7}$, K.~Hurley$^8$, I.~Arcavi$^9$, S.B.~Cenko$^{10,11}$, A.~Gal-Yam$^9$, A.~Horesh$^{12}$, M.~Kasliwal$^{13,14}$, D.~Poznanski$^{15}$, J.M.~Silverman$^{16,17}$, M.~Sullivan$^{18}$, J.S.~Bloom$^{11}$, A.V.~Filippenko$^{11}$, S.R.~Kulkarni$^{12}$, P.E.~Nugent$^{11,19}$, E.~Ofek$^9$, S.~Barthelmy$^{10}$, W.~Boynton$^{20}$, J.~Goldsten$^{21}$, S.~Golenetskii$^{22}$, M.~Ohno$^{23}$, M.S.~Tashiro$^{24}$, K.~Yamaoka$^{25}$, X.L-.~Zhang$^{26}$} 
\vspace{0.4cm}\\
\parbox{\textwidth} 
{$^1$Scuola Normale Superiore di Pisa, Piazza dei Cavalieri 7, 56126 Pisa, Italy\\
$^2$Yale University, Department of Physics, P.O. Box 208120, New Haven, CT 06520-8120, USA\\
$^3$Liverpool John Moores University, Astrophysics Research Institute, Liverpool Science Park, IC2 Building, 146, Brownlow Hill, Liverpool, L3 5RF, UK\\
$^4$INAF-Padova Astronomical Observatory, Vicolo dell'Osservatorio 5, 35122, Padova, Italy\\
$^5$Max-Planck Institute for Astrophysics, Garching, Karl-Schwarzschild-Str. 1, Postfach 1317, D-85741 Garching, Germany\\
$^6$INAF-IASF, Via P. Gobetti 101, 40129 Bologna, Italy\\
$^7$INFN, Sezione di Pisa, Largo Pontecorvo 3, 56127 Pisa, Italy\\
$^8$U.C. Berkeley Space Sciences Laboratory, 7 Gauss Way, Berkeley, CA 94720-7450, U.S.A.\\
$^9$ Department of Particle Physics and Astrophysics, The Weizmann Institute of Science, Rehovot 76100, Israel \\
$^{10}$ NASA Goddard Space Flight Center, Mail Code 661, Greenbelt, MD 20771, USA \\
$^{11}$ Department of Astronomy, University of California, Berkeley, CA 94720-3411, USA\\
$^{12}$Cahill Center for Astrophysics, California Institute of Technology, Pasadena, CA 91125, USA\\
$^{13}$ Hubble Fellow \\
$^{14}$ Carnegie/Princeton Fellow, 813 Santa Barbara St, Pasadena, CA 91101, USA\\
$^{15}$ School of Physics and Astronomy, Tel-Aviv University, Tel-Aviv 69978, Israel\\
$^{16}$ Department of Astronomy, University of Texas at Austin, Austin, TX 78712, USA\\
$^{17}$ NSF Astronomy and Astrophysics Postdoctoral Fellow\\
$^{18}$ School of Physics and Astronomy, University of Southampton, Southampton, SO17 1BJ, UK\\
$^{19}$ Lawrence Berkeley National Laboratory, 1 Cyclotron Road, Berkeley, CA 94720, USA\\
$^{20}$ University of Arizona, Lunar and Planetary Laboratory, Tucson, AZ 85721, U.S.A.\\
$^{21}$ Applied Physics Laboratory, Johns Hopkins University, Laurel, MD 20723, U.S.A.\\
$^{22}$ Ioffe Physico-Technical Institute of the Russian Academy of Sciences, St. Petersburg, 194021, Russian Federation\\
$^{23}$ Department of Physical Science, Hiroshima University, 1-3-1 Kagamiyama, Higashi-Hiroshima, Hiroshima 739-8526, Japan\\
$^{24}$ Department of Physics, Saitama University, 255 Shimo-Okubo, Sakura-ku, Saitama 338-8570, Japan\\
$^{25}$ Department of Physics and Mathematics, Aoyama Gakuin University, 5-10-1 Fuchinobe, Sagamihara, Kanagawa 229-8558, Japan \\
$^{26}$ Max-Planck-Institut f\"{u}r extraterrestrische Physik, Giessenbachstrasse, Garching, 85748 Germany\\
}}

\begin{document}

\date{Released  XXXXX}

\pagerange{\pageref{firstpage}--\pageref{lastpage}} 
\pubyear{2013}

\maketitle 

\label{firstpage}
\begin{abstract}
	We present optical photometry and spectroscopy of the broad-lined Type Ic supernova (SN~Ic-BL) PTF10qts, which was discovered as part of the Palomar Transient Factory. The supernova was located in a dwarf galaxy of magnitude $r=21.1$ at a redshift $z=0.0907$. We find that the $R$-band light curve is a poor proxy for bolometric data and use photometric and spectroscopic data to construct and constrain the bolometric light curve.  The derived bolometric magnitude at maximum light is $M_{\rm bol} = -18.51\pm0.2$ mag, comparable to that of SN\,1998bw ($M_{\rm bol} = -18.7$ mag) which was associated with a gamma-ray burst (GRB). PTF10qts is one of the most luminous SN~Ic-BL observed without an accompanying GRB. We estimate the physical parameters of the explosion using data from our programme of follow-up observations, finding that it produced a larger mass of radioactive nickel compared to other SNeIc-BL with similar inferred ejecta masses and kinetic energies. The progenitor of the event was likely a $\sim20$\,M$_{\odot}$ star.  
\end{abstract}

%
\section{Introduction}\label{sec:introduction}

Type Ic supernovae (SNe~Ic) are classified from their optical spectra as having no hydrogen or helium present \citep[for a review of supernova classification, see][]{F97}. They constitute $\sim 10$\% of the total number of SNe in the local universe \citep{Li:2011cl}. SNe~Ic are believed to be core-collapse events from either a massive Wolf-Rayet (WR) star that has lost its outer layers via a wind-loss mechanism \citep{Gaskell:1986ge}, or a less massive star where the envelope has been stripped by a binary companion \citep{Podsiadlowski:1992ij,Nomoto:1995ej}. For a recent review of the progenitors of all core-collapse SNe, see \citet{Smartt:2009kr}.

One subgroup of SNe~Ic, referred to as broad-lined Type Ic supernovae (SNe~Ic-BL) or sometimes ``hypernovae,'' exhibits very high line velocities in the spectra, indicating an explosion with high kinetic energy per unit mass. These objects have been linked to gamma-ray bursts (GRBs), initially with the observation that the broad-lined, energetic SN\,1998bw was coincident with the long-duration GRB\,980425 \citep{Galama:1998ea}. Subsequently, five other spectroscopically confirmed SNe have been identified with GRBs or X-ray flashes (XRFs) between redshifts $z$ of 0.03 and 0.2: GRN\,030329/SN\,2003dh \citep{Hjorth:2003jv,Stanek:2003ef,Matheson:2003ey}, GRB\,031203/SN\,2003lw \citep{Malesani:2004do,GalYam:2004hh,Thomsen:2004fq,Cobb:2004jj}, XRF\,060218/SN\,2006aj \citep{Pian:2006ho,Mirabal:2006fg,Sollerman:2006bv,Modjaz:2006dl,Cobb:2006jy,Ferrero:2006dm}, XRF\,100316D/SN\,2010bh \citep{Starling:2011jm,Chornock:2010ue,Cano:2011jl,Bufano:2012ke}, and GRB\,130702A/SN\,2013dx \citep{Schulze:2013grb,Cenko:2013grb,DElia:2013grb,Singer:2013aa}. There are also a large and growing number of cases where the optical afterglow of GRBs or XRFs exhibit features typical of (or consistent with) those of SNe~Ic-BL \citep[for example,][]{Soderberg:2005jb,Bersier:2006gu,Cano:2011cy,Berger:2011ee,Sparre:2011go,Melandri:2012ce,Xu:2013ww,Levan:2013vg,Jin:2013cf}.

However, there are also many examples of high-energy SNe~Ic for which no associated GRB has been found, including SN\,1997ef \citep{Mazzali:2000cx} and SN\,2002ap \citep{Mazzali:2002bf,GalYam:2002fc,Foley:2003hp}. It has been suggested that all high-energy SNe~Ic form GRBs and that we do not observe the gamma-ray jet because of our viewing angle \citep{Podsiadlowski:2004dn}. This hypothesis is supported by the rates and measurements of the energetics \citep{Smartt:2009ge}, but not by radio observations \citep{Soderberg:2006hq}.

The Palomar Transient Factory \citep[PTF;][]{Law:2009cq,Rau:2009fp} was an optical survey of the variable sky using a 7.3 square degree camera installed on the 48-inch Samuel Oschin telescope at Palomar Observatory. PTF conducted real-time analysis and had a number of follow-up programmes designed to obtain colours and light curves of detected transients from a variety of facilities \citep{GalYam:2011fv}. A major science goal of PTF was to conduct a SN survey free from host-galaxy bias and sensitive to events in low-luminosity hosts. Such a survey was particularly suitable to search for SNe~Ic-BL, which appear to be more abundant in low-luminosity dwarf galaxies \citep{Arcavi:2010ky}.

In this paper, we present optical photometry and spectra of PTF10qts, a SN~Ic-BL. Section \ref{sec:observations} describes the observations, which are analysed in Section \ref{sec:discuss}. We summarise our results in Section \ref{sec:conc}. Throughout the paper, we assume $\Omega_M = 0.3$, $\Omega_{\Lambda} = 0.7$, and H$_0 = 70$\,km\,s$^{-1}$\,Mpc$^{-1}$.

\begin{figure}
	\centering 
	\includegraphics[width=\columnwidth]{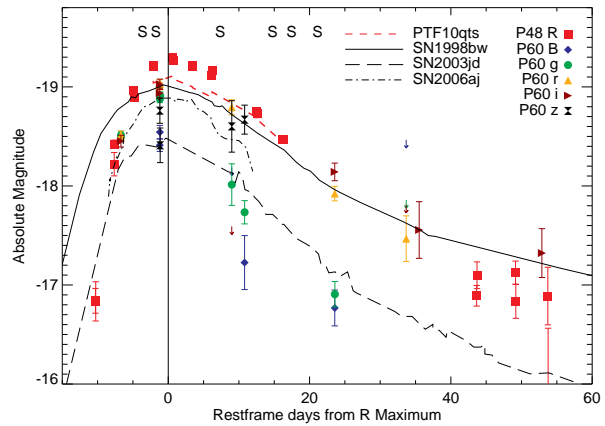} \caption{Light curve of PTF10qts from photometry taken on the P48 and P60 telescopes. The data points are given in Table \ref{tab:phot} and converted to absolute magnitudes. The solid line represents the $R$-band light curve of  SN\,1998bw \citep{Galama:1999et}, the first GRB-SN, and the dot-dashed line shows the $R$-band light curve of SN\,2006aj \citep{Ferrero:2006dm}, which was accompanied by an X-ray flash. The dashed line shows SN\,2003jd \citep{Valenti:2008dz}.  No K-corrections have been applied to the individual PTF10qts data, except the long-dashed red line which shows the PTF10qts $R_{\rm PTF}$ points K-corrected using spectra where possible.} \label{fig:lc_all} 
\end{figure}

\begin{table}
	\caption{$R_{\rm PTF}$ photometry taken with the P48 and $Bgriz$ with the P60. Dates and phases are given in the rest frame relative to $R$-band maximum, and the photometry has been corrected for Galactic extinction ($E(B-V)=0.029$\,mag).  These data are plotted in Figure \ref{fig:lc_all} after conversion to absolute magnitudes assuming a distance modulus of $\mu = 415$\,mags.  The first line of the table is the upper limit of the last non-detection before the supernova was discovered. \label{tab:phot} }
	\begin{center}
		\begin{tabular}
			{c c c c c} \hline MJD & Band & Phase & Magnitude & Error \\
			&&(days)\\
			\hline 
			   55410.288 &  $R_{\rm PTF}$  & -13.03 & $>$20.1\\
			   55413.260 &  $R_{\rm PTF}$  & -10.31 &  21.25 &  0.13\\
		       55413.304 &  $R_{\rm PTF}$  & -10.26 &  21.26 &  0.20\\
		       55416.160 &  $R_{\rm PTF}$  & -7.65 &  19.87 &  0.12\\
		       55416.204 &  $R_{\rm PTF}$  & -7.61 &  19.67 &  0.04\\
		       55419.175 &  $R_{\rm PTF}$  & -4.88 &  19.13 &  0.03\\
		       55419.218 &  $R_{\rm PTF}$  & -4.84 &  19.19 &  0.02\\
		       55422.170 &  $R_{\rm PTF}$  & -2.14 &  18.87 &  0.03\\
		       55422.214 &  $R_{\rm PTF}$  & -2.10 &  18.89 &  0.02\\
		       55425.200 &  $R_{\rm PTF}$  & 0.64 &  18.79 &  0.03\\
		       55425.242 &  $R_{\rm PTF}$  & 0.68 &  18.82 &  0.02\\
		       55428.199 &  $R_{\rm PTF}$  & 3.39 &  18.87 &  0.02\\
		       55428.252 &  $R_{\rm PTF}$  & 3.44 &  18.88 &  0.02\\
		       55431.236 &  $R_{\rm PTF}$  & 6.18 &  18.98 &  0.02\\
		       55431.280 &  $R_{\rm PTF}$  & 6.22 &  18.92 &  0.03\\
		       55438.160 &  $R_{\rm PTF}$  & 12.52 &  19.34 &  0.04\\
		       55438.204 &  $R_{\rm PTF}$  & 12.56 &  19.36 &  0.05\\
		       55442.192 &  $R_{\rm PTF}$  & 16.22 &  19.62 &  0.03\\
		       55442.240 &  $R_{\rm PTF}$  & 16.26 &  19.62 &  0.03\\
		       55472.110 &  $R_{\rm PTF}$  & 43.65 &  21.20 &  0.10\\
		       55472.162 &  $R_{\rm PTF}$  & 43.70 &  20.99 &  0.14\\
		       55478.111 &  $R_{\rm PTF}$  & 49.15 &  20.97 &  0.12\\
		       55478.154 &  $R_{\rm PTF}$  & 49.19 &  21.26 &  0.17\\
		       55483.097 &  $R_{\rm PTF}$  & 53.72 &  21.20 &  0.29\\
		       55483.141 &  $R_{\rm PTF}$  & 53.76 &  22.19 &  0.66\\
			
	\hline
	   55423.197 &  $B$ &  -1.19 &  19.68 &  0.06\\
       55423.211 &  $B$ &  -1.18 &  19.51 &  0.07\\
       55436.286 &  $B$ &  10.81 &  20.79 &  0.27\\
       55450.204 &  $B$ &  23.57 &  21.25 &  0.18\\
       \hline
	55417.201 &  $g$ &  -6.69 &  19.59 &  0.04\\
        55423.206 &  $g$ &  -1.19 &  19.15 &  0.04\\
       55423.259 &  $g$ &  -1.14 &  19.20 &  0.04\\
       55434.307 &  $g$ &  8.99 &  20.02 &  0.21\\
       55436.289 &  $g$ &  10.81 &  20.37 &  0.12\\
       5450.220 &  $g$ &  23.58 &  21.12 &  0.13\\
       55506.082 &  $g$ &  74.80 &  21.65 &  0.15\\
       55515.090 &  $g$ &  83.06 &  21.60 &  0.35\\
\hline
       55423.195 &  $r$ &  -1.20 &  19.05 &  0.04\\
       55423.210 &  $r$ &  -1.18 &  19.03 &  0.04\\
       55434.302 &  $r$ &  8.99 &  19.34 &  0.08\\
        55450.203 &  $r$ &  23.57 &  20.13 &  0.07\\
       55461.229 &  $r$ &  33.67 &  20.66 &  0.23\\
       55513.092 &  $r$ &  81.22 &  21.38 &  0.23\\
       55515.086 &  $r$ &  83.05 &  21.06 &  0.19\\
\hline
       55417.198 &  $i$ &  -6.69 &  19.70 &  0.06\\
       55423.193 &  $i$ &  -1.20 &  19.14 &  0.05\\
       55423.208 &  $i$ &  -1.18 &  19.12 &  0.05\\
       55450.201 &  $i$ &  23.56 &  19.93 &  0.09\\
       55463.233 &  $i$ &  35.51 &  20.59 &  0.29\\
       55482.180 &  $i$ &  52.88 &  20.75 &  0.25\\
       55506.076 &  $i$ &  74.79 &  21.42 &  0.31\\
       55513.090 &  $i$ &  81.22 &  21.00 &  0.24\\
       55515.085 &  $i$ &  83.05 &  21.42 &  0.35\\
\hline
       55423.205 &  $z$ &  -1.19 &  19.33 &  0.13\\
       55423.251 &  $z$ &  -1.15 &  19.75 &  0.17\\
       55434.305 &  $z$ &  8.99 &  19.49 &  0.26\\
       55436.287 &  $z$ &  10.81 &  19.42 &  0.15\\

			\hline 
		\end{tabular}
	\end{center}
\end{table}

\begin{figure}
	\begin{center}
		\includegraphics[width=\columnwidth]{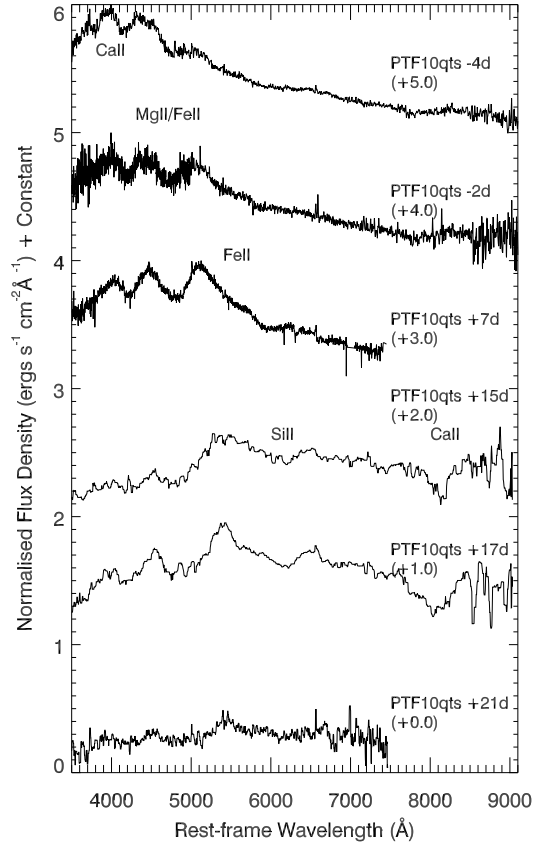} \caption{Photospheric spectra of PTF10qts. The phases are given in the rest frame relative to $R$-band maximum. The spectra are labelled showing important species discussed in the text.  The number in brackets shows the offset applied to each spectrum.} \label{fig:spec-all} 
	\end{center}
\end{figure}

\section{Observations}\label{sec:observations}

\subsection{Optical Photometry}\label{sec:phot}

On 2010 August 05.230 (UT dates are used throughout this paper), PTF10qts was discovered by the Palomar 48-inch telescope \citep[P48,][]{Rahmer:2008hs} at $R_{\rm PTF} \approx 20.3$\,mag\footnote{$\lambda_c$ = 6540\AA}; its coordinates were $\alpha$(J2000) = 16$^h$41$^m$37.60$^s$, $\delta$(J2000) = +28$^\circ$58$'$21.1$''$. It was also detected again in an image taken later that night on 2010 August 05.305. It was not detected in an image taken three nights previously with a limit of $R_{\rm PTF} = 20.1$\,mag. A small, faint ($r = 21.1$\,mag) host galaxy, object J164137.53+285820.3, is visible in the SDSS catalogue 1.2" away from the supernova. Its redshift, $z=0.0907$, was measured from H$\alpha$ and [\ion{O}{ii}] narrow emission lines in spectra of PTF10qts. At the host's distance (414.9\,Mpc), the supernova offset from the centre of this corresponds to a projected physical distance of 2.4\,kpc. No host is discernible in the P48 images. PTF10qts was also observed at the Palomar 60-inch telescope \citep[P60;][]{Cenko:2006im} in $B$, $g$, $r$, $i$, and $z$, although the cadence for these observations was less than at the P48. 

For observations with the P48, measurements were performed by standard image subtraction, using a deep, good-seeing reference constructed from images taken before the SN exploded. The reference was matched astrometrically to field stars in each image containing the SN and subtracted, and point-spread function (PSF) photometry was then performed. For the P60 data, we employed direct aperture photometry without host-galaxy subtraction, as the host is very faint. The data are calibrated to SDSS magnitudes. A light curve is plotted in Figure \ref{fig:lc_all} and listed in Table \ref{tab:phot}. We give the phase in the rest frame relative to $R$-band maximum determined from fitting the points around maximum light with a parabola.  The MJD of the $R$-band maximum is $55424.6 \pm 0.5$ (16.6 August 2010).  With the non-detection on 2 August 2010 (MJD = 55410.244), we can determine the date of explosion to within 3 days; thus, we constrain the rise time in the $R$ band to $12.7 \pm 1.5$\,days in the observed frame or $11.6 \pm 1.4$\,days in the rest frame.

The data points in Figure \ref{fig:lc_all} have been corrected for Milky Way extinction, $E(B-V) = 0.029$\,mag, using the dust maps of \citet[][]{Schlegel:1998fw} and the extinction curve of \citet{Cardelli:1989dp}. The equivalent width of the \ion{Na}{i}~D line at zero redshift measured in the spectrum of PTF10qts taken at +7\,days is $0.15\pm0.12$\,\AA, which can be converted to a measurement of extinction via the relation of \citet{Turatto:2003}. The measured value of $E(B-V) = 0.024\pm0.019$\,mag is consistent with that determined from the dust maps (but see \citealt{Poznanski:2011cc}). This is also consistent with the value derived from \citet{Poznanski:2012bf} of $E(B-V) = 0.021^{+0.09}_{-0.014}$\,mag.  We observe no \ion{Na}{i}~D at the redshift of the SN, so no correction has been applied for host-galaxy extinction.  The correction for Milky Way extinction has been applied to the individual points shown in Figure \ref{fig:lc_all}, with no K-corrections (see below).

Figure \ref{fig:lc_all} also shows the $R$-band light curves of three other SNe~Ic for comparison. SN\,1998bw is a broad-lined SN~Ic and the first GRB-SN; the values of $M_R$ are similar for both objects. We also include SN\,2006aj, which was accompanied by an X-ray flash, and SN\,2003jd, which appears to be spectroscopically similar to PTF10qts (see Section \ref{sec:spec}). From the raw $R$-band light curve, it appears that the SN reaches a more luminous absolute magnitude than SN\,1998bw, but with a light-curve width more similar to those of SN\,2003jd and SN\,2006aj.  The long-dashed line shows the $R$-band light curve of PTF10qts with K-corrections based on the photospheric spectra. This confirms that the $R$-band light curve is slightly more luminous than that of SN\,1998bw, but the decline rate is faster.  K-corrections are discussed in more detail in Section \ref{sec:rband}.

\begin{figure}
	\begin{center}
		\includegraphics[width=\columnwidth]{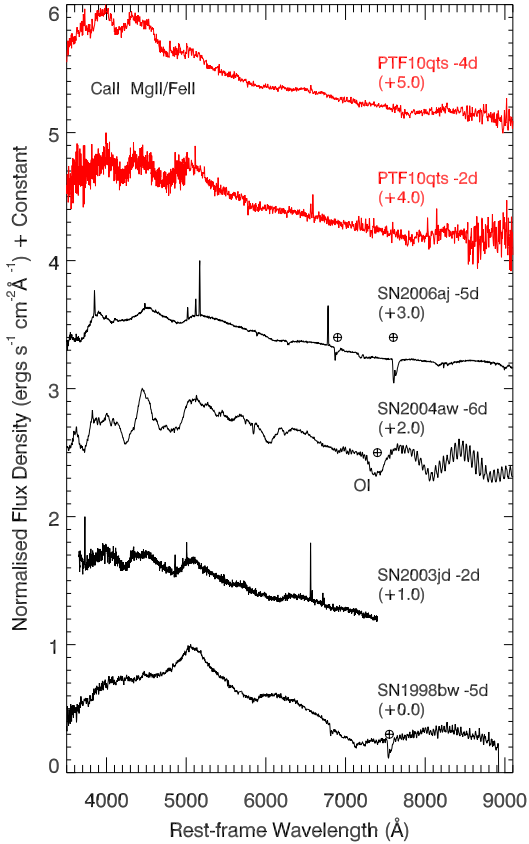} 
		\caption{A comparison of spectra before maximum light. Telluric features are marked.  The spectra are labelled showing important species discussed in the text.  The number in brackets shows the offset applied to each spectrum.}
		 \label{fig:specpremax}
	\end{center}
\end{figure}

\begin{figure}
	\begin{center}
		\includegraphics[width=\columnwidth]{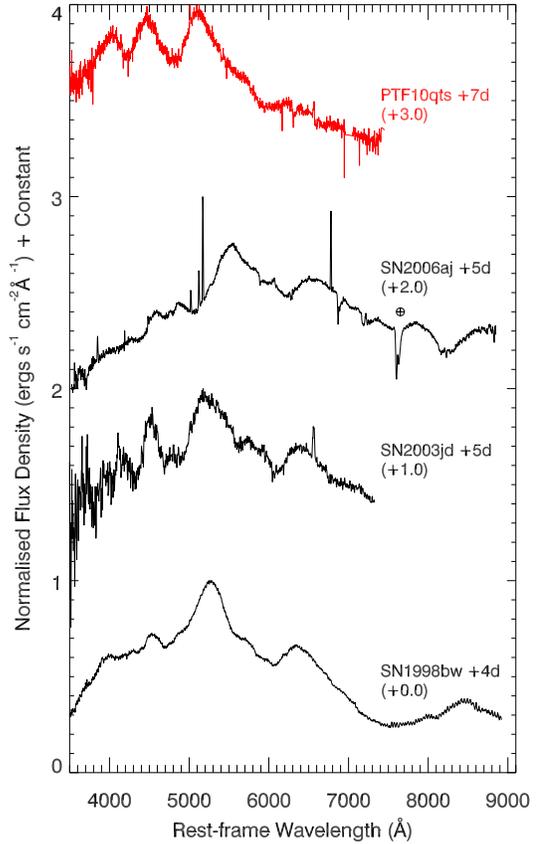} \caption{A comparison of spectra around +7 days after maximum light. 
Telluric features are marked.  The number in brackets shows the offset applied to each spectrum.} 
		\label{fig:spec-7} 
	\end{center}
\end{figure}
\subsection{Optical Spectroscopy}\label{sec:spec}

\begin{table*}
	\centering \caption{A summary of spectroscopic observations of PTF10qts.  The phase is relative to $R$-band maximum (16.6 August 2010) and then converted to the rest frame.  Note that for the Lick/Kast spectrum, the blue side resolution is 4.1\AA\ (FWHM) and the red side resolution is 9.1\AA\ (FWHM).  The velocities, as plotted in Figure \ref{fig:velocities} are determined from the \ion{Si}{ii} 6355\AA\ line.} \label{tab:spec-obs} 
	\begin{tabular}
		{c c c c c c c c } \hline Date & Phase &  Telescope &  Range & Resolution & Velocity \\
		(UT) & (days) & &(\AA) & (\AA\,pix$^{-1}$) & (1000 km/s)\\
		\hline 2010-08-13 & -3.6 &P200/DBSP & 3505--10,100 & 5 & --\\
		2010-08-15 & -1.8 & Lick/Kast &  3480--10,000 & 4.1/9.1 & 19.1 $\pm$ 0.75\\
		2010-08-25 & +7.4 & TNG/DOLORES & 3360--8050 & 2.25 &14.4 $\pm$ 0.5\\
		2010-09-02 & +14.8 & P200/DBSP & 3440--9850 & 2 & 12.0 $\pm$ 0.5\\
		2010-09-05 & +17.4 & P200/DBSP &  3440--9850 & 2 & 8.5 $\pm$ 0.75\\
		2010-09-09 & +21.2 &  KPNO/RC Spec &  3620--8140 & 5.5& --\\
		2011-04-27 & +231.7 & Keck/LRIS &  3100--10,200 & 2 & --\\
		\hline 
	\end{tabular}
\end{table*}

Follow-up spectroscopy of PTF10qts was carried out at a number of international observatories and is summarised in Table \ref{tab:spec-obs}. The SN was classified as an SN~Ic-BL based on its broad features and lack of obvious hydrogen and helium, and weak silicon in the spectra. The photospheric spectra are plotted in Figure \ref{fig:spec-all}, where all phases are given relative to $R$-band maximum for each object. Standard \textsc{IRAF} routines as well as custom \textsc{IDL} procedures were used to remove bias and flat field correct the spectra, as well as to create wavelength and flux solutions for the data. These were then applied to the frames and calibrated spectra were extracted from the data.  All spectra of PTF10qts are publically available via WISeREP\footnote{http://www.weizmann.ac.il/astrophysics/wiserep/} \citep{Yaron:2012bp}.

\begin{figure}
	\begin{center}
		\includegraphics[width=\columnwidth]{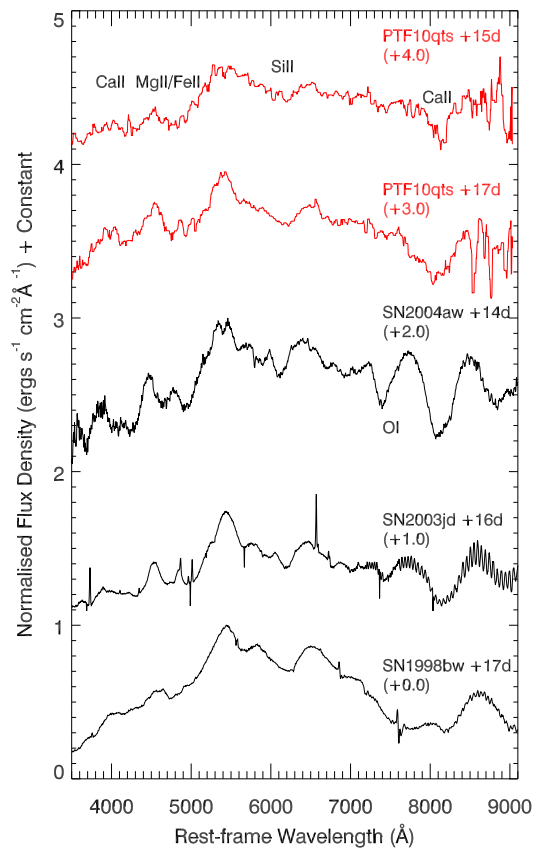} 
		\caption{A comparison of spectra around +14 days after maximum light.  The spectra are labelled showing important species discussed in the text.  The number in brackets shows the offset applied to each spectrum.} \label{fig:spec-14} 
	\end{center}
\end{figure}

\begin{figure}
	\begin{center}
		\includegraphics[width=\columnwidth]{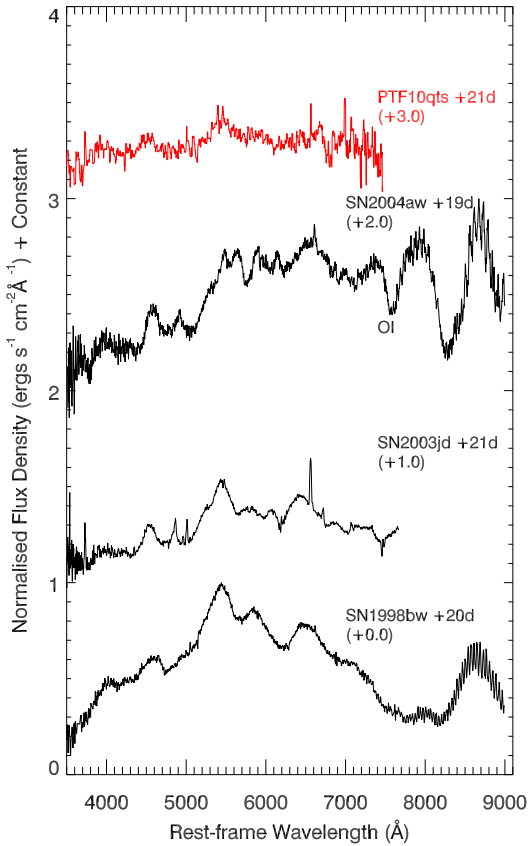} 
		\caption{A comparison of spectra around +21 days after maximum light.  The spectra are labelled showing important species discussed in the text.  The number in brackets shows the offset applied to each spectrum.} \label{fig:spec-21} 
	\end{center}
\end{figure}

We compare the spectra of PTF10qts to those of other known SNe~Ic and SNe~Ic-BL  in Figures \ref{fig:specpremax}--\ref{fig:spec-21}. Again, all phases are given relative to the $R$-band maximum for that particular object.  We divide our spectra into four periods of observation --- pre-maximum, $+7$\,days after $R$-band maximum, $+14$\,days, and $+21$\,days --- consider each of these separately.

Before maximum light, the spectrum of PTF10qts is dominated by broad, high-velocity absorption lines which are blended together.  The absorptions at 4400\,\AA\ and 4800\,\AA\ are dominated by \ion{Fe}{ii}.  \ion{Si}{ii} may be seen in the $-2$\,days spectrum and later as the elbow at 5800\,\AA, but it is blended with other features, making isolation of this feature and a measurement of the photospheric velocity difficult.  We also note that visually, the features at 4000--6000\,\AA\ of SN\,2006aj \citep{Pian:2006ho} are similar to the early phases of PTF10qts: the spectrum is blue and contains broad absorptions around 4000\,\AA. We do not see the absorption due to \ion{O}{i} in the 7000--7600\,\AA\ region visible in spectra of SN\,1998bw \citep{Patat:2001jt} or SN\,2004aw \citep{Taubenberger:2006cc}, another SN~Ic-BL .  The $t = -2$\,day spectrum is redder than the $t=-4$\,day spectrum, reflecting the fact that the temperature is decreasing as the ejecta expand.

Around a week past maximum brightness, the spectrum of PTF10qts resembles that of SN\,2003jd \citep{Valenti:2008dz}, both in the three broad absorption features in the blue and the shape of the continuum in the red.

Our next phase of spectroscopy is around two weeks past maximum light. As seen in Figure \ref{fig:spec-14}, the SN ejecta have expanded sufficiently that we can now see individual features, such as the \ion{Si}{ii} lines around 6200\,\AA\ and strong \ion{Ca}{ii} absorption at 8100\,\AA. The velocity of the \ion{Ca}{ii} near-infrared triplet is $\sim$\,18,000\,km\,s$^{-1}$ (measured from the blueshift of the feature's minimum), which is higher than for both SN\,2003jd and SN\,2004aw as shown in the figure. In the blue we also see absorption caused by \ion{Mg}{ii}, \ion{Ca}{ii}, and \ion{Fe}{ii}. Visually the spectra retain their similarity to those of SN\,2003jd, SN\,2004aw, and to a lesser degree SN\,1998bw without \ion{O}{i} which is still absent.  This may be indicative of a smaller ejecta mass.

The final spectrum of PTF10qts taken during the photospheric phase is shown in Figure \ref{fig:spec-21}. It is quite noisy, but visually the spectral evolution continues to be similar to those of SN\,2003jd and SN\,2004aw.  There may be slight absorption from \ion{O}{i} visible in these later spectra.  This raises the possibility of a sequence of oxygen masses in SNe Ic-BL ranging from strong in supernovae such as SN2004aw, through objects like SN2003jd and finally objects like PTF10qts which show no oxygen.

From this spectral comparison, we conclude that PTF10qts is not a good match to any single well-observed SN~Ic-BL over its entire evolution, although at some phases there appear to be reasonable matches to other known SNe~Ic-BL.  PTF10qts lacks the very high velocities (i.e., energy per unit mass) of SN\,1998bw, and the spectral features (related to element abundances) do not match those seen in the lower-velocity examples of SN\,2003jd, SN\,2004aw, and SN\,2006aj.

\begin{figure}
	\begin{center}
		\includegraphics[width=\columnwidth]{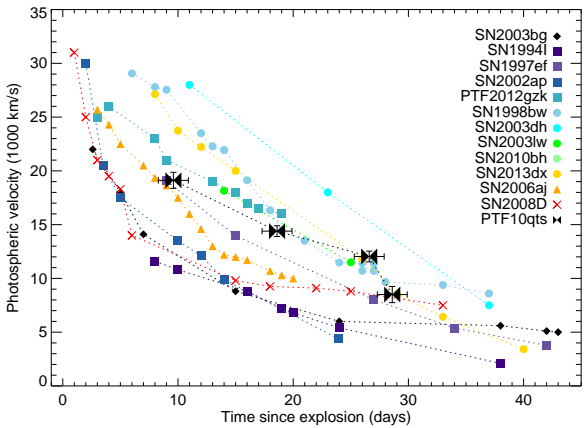} 
		\caption{A plot of the photospheric velocities of supernovae at different times after explosion.  Different symbol shapes correspond to different types of supernova: diamond - IIb; square - Ic/Ic-BL; circle - GRB/SNe; triangle - XRF/SN Ic; cross - XRF/SN Ib; and bowtie - PTF10qts.  This figure is augmented from the one produced in \citet{Mazzali:2008vn}.} \label{fig:velocities} 
	\end{center}
\end{figure}

\section{Discussion}\label{sec:discuss}

\subsection{Velocity Determination}\label{sec:vel}

Before determining the physical parameters of this supernova, it is necessary to constrain the supernova photospheric velocity at maximum light, which is typically characterised by the minimum of the blueshifted absortion of the \ion{Si}{ii} feature around $\sim6150$\AA.  However, as seen below, we use two different methods which involve two different dates for maximum light and we do not have the spectral coverage around these times to measure values directly from spectra, along with the additional problem of the supernova features being blended together so the \ion{Si}{ii} line which is thought to give a clear determination of the photospheric velocity is not always visible as a separate feature.

In Figure \ref{fig:velocities}, we show the velocities of PTF10qts (bowtie) compared to a number of other types of supernovae, including the two we will use for analogues - SN1998bw and SN2006aj.  We can define the velocity to use in later analysis in different ways such as the velocity at maximum light in $R-band$ or the velocity at the maximum of the bolometric light curve.  The time between explosion and maximum varies between supernovae, but if we wanted a uniform time, we could also take a fixed date after explosion.

Applying these three methods to PTF10qts reveals that both the $R$-band and bolometric maxima fall between the first two velocity measurements from the spectra.  We interpolate linearlly between the velocities measured at $-1.8$\,days and $+7.4$\,days as given in Table \ref{tab:spec-obs} and assign PTF10qts a photospheric velocity of 17000$\pm$1500kms$^{-1}$.  Due to the small difference in rise times for the bolometric and $R$-band light curves for SN1998bw and SN2006aj we use two different velocities for these objects in the following sections and assign an error of $\pm$1000kms$^{-1}$ to each one.  We have been conservative with the velocity errors, but they are only a small contribution to the final error on the physical parameters we derive below.

\subsection{$R$ Band as a Proxy for Bolometric}\label{sec:rband}

We first attempt to use the $R$-band light curve, which has the best phase coverage, to estimate some of the physical parameters of the SN explosion.  It has also been suggested that $R$-band can be used as a proxy for bolometric when calculating the physical parameters of the explosion \citet{Drout:2011iw}.  We employ a well-studied example as an analogue and scale the physical parameters based on the modeling of that object.  Ideally, the analogue would match the light curve and the spectrum.  This is particularly important for SNe~Ic-BL, as the kinetic energy is dominated by the broadest parts of the lines.  These features are usually blended; thus, as well as matching the velocities, it is important to match the spectra to reduce the error when scaling the parameters.

Unfortunately, as discussed above, there is no single good analogue of PTF10qts. Instead, we take two examples for which bolometric light curves and other physical parameters are well modeled, and we use their properties to estimate the ejecta mass $M_{\rm ej}$, kinetic energy $E_{\rm K}$, and nickel mass $M(^{56}\textrm{Ni})$ of PTF10qts. We adopt SN\,1998bw, as this is the most similar in absolute $R$ magnitude to PTF10qts at maximum light ($R=-19.16$\,mags for SN\,1998bw compared to $R=_{\rm PTF}=-19.30$\,mags for PTF10qts), and also SN\,2006aj, which is similar spectroscopically. Radiative transfer models of their light curves have been developed to derive the physical parameters of these explosions \citep{Nakamura:2001gw,Mazzali:2006bm}.	

In order to compare properly PTF10qts with existing samples in the literature, we use photometry and spectra to measure $R-I$ colours and obtain the transformation from magnitudes in $R_{\rm PTF}$ to $R$ as in \citet{Ofek:2012aa} and \citet{Jordi:2006do}. We find that given the sparse light-curve coverage, it is not possible to infer a relationship between $R-I$ and phase. We therefore assume a constant value of $R-I = 0.24\pm0.12$\,mag, which is the mean of all the measured values. This corresponds to $R = R_{\rm PTF} - 0.14\pm0.01$\,mag, where the quoted uncertainty comes only from the colour term.

Given the redshift of PTF10qts, the observed $R$ band is very different from the observed $R$ band of the local comparison SNe we have used.  To compensate for this, we calculate K-corrections using the spectra of PTF10qts following \citet{Humason:1956gc}.  The spectra acquired on 25 August (TNG) and 9 September (KPNO) fall short of covering the full $R$ band by a few hundred Angstroms when shifted to the rest frame.  Therefore, at wavelengths longer than their red end, we assumed that their behaviour is similar to the  spectra taken on 15 August (Lick) and 2 September (P200), respectively, based on the similarity of the spectra at bluer wavelengths.  We then interpolated the measurements to obtain K-corrections at 0 and +15 days relative to $R$-band maximum.  The calculated values are $-0.174$ and $-0.027$, respectively.

We interpolated the $R$-band light curve of PTF10qts to obtain final values of $R(0) = -19.27\pm0.06$ and $R(15) = -18.69\pm0.06$\,mag. The uncertainties include measurement errors, uncertainties in the K-correction, and conversion from $R_{\rm PTF}$ to $R$. We therefore find $\Delta m_{15}(R) = 0.58\pm0.08$\,mag for PTF10qts.  This is similar to that of SN\,1998bw ($\Delta m_{15}(R) = 0.56$\,mag), but much smaller than that of SN\,2006aj ($\Delta m_{15}(R) = 0.86$\,mag).  Note that the K-corrected values in $R$ differ from the light curve for PTF10qts shown in Figure \ref{fig:lc_all}, as that is for $R_{\rm PTF}$.

Using \cite{Arnett:1982cv}, we know the following relations for a SN at maximum light:
\begin{eqnarray}
	M_{\rm ej} &\propto & \tau^2 v_{\rm phot},~{\rm and} \label{eq:mass} \\
	E_{\rm K} & \propto& \tau ^2 v_{\rm phot}^3, \label{eq:energy} 
\end{eqnarray}

\noindent where $\tau$ is the light-curve width which is proportional to $1/\Delta m_{15}(R)$, and $v_{\rm phot}$ is the photospheric velocity. We have chosen to use $\Delta m_{15}(R)$ instead of $\tau$ for this measurement because of the uncertainty in the K-corrections before the first epoch of spectroscopy.  For $v_{\rm phot}$ we adopt the values of 15,500\,km\,s$^{-1}$ for SN\,2006aj and 18,000\,km\,s$^{-1}$ for SN\,1998bw. As discussed above, we have assumed $v_{\rm phot} =$ 17,000\,km\,s$^{-1}$. With Equations \ref{eq:mass} and \ref{eq:energy}, we can calculate the physical parameters for PTF10qts assuming that it is analogous to either SN\,1998bw or SN\,2006aj\footnote{Note that for SN\,2006aj, the light-curve data for the $R$ band ends at +12 days relative to $R$-band maximum, but the bolometric light curve extends to +14 days owing to the availability of data in other filters.  We evaluated bolometric magnitudes from these by assuming a constant bolometric correction with respect to the $V$ band.  We obtain the same result if we extrapolate just the $R$-band light curve over the longer interval.}. The resulting parameters are given in Table \ref{tab:phys-params}.

\begin{table*}
	\centering \caption{A summary of the physical parameters assumed for SN\,1998bw and SN\,2006aj, and those derived for PTF10qts first in the $R$ band and then using the bolometric light curve. $E_{\rm K}$ is the kinetic energy and $L_{i}$ is the peak luminosity in either $R$ or bolometric. The parameters for SN\,1998bw and SN\,2006aj are either measured from their published light curves or, in the case of $E_{\rm K}$, $v$, $M_{\rm ej}$, and $M(^{56}\textrm{Ni})$, taken from the modeling in \citet{Nakamura:2001gw} and \citet{Mazzali:2006bm} respectively.} 
	\label{tab:phys-params} 
	\begin{tabular}
		{c c c c c}
		
		\hline Parameter & SN\,1998bw & SN\,2006aj & PTF10qts& PTF10qts\\
		&&& SN\,1998bw-like & SN\,2006aj-like \\
		\hline 
		\multicolumn{5}{c}{$R$ band}\\
		\hline
		$\Delta m_{15}(R)$ (mag)& 0.56 & 0.86 &0.58$\pm$0.18 & 0.58$\pm$0.18 \\
		$v$ (km\,s$^{-1}$)& 19,000$\pm$1000 &15,000$\pm$1000 &\multicolumn{2}{c}{17,000$\pm$1500}\\
		$L_{R}$ (ergs\,s$^{-1}$) & 1.87 $\times 10^{42}$& 1.08$\times 10^{42} $& \multicolumn{2}{c}{(2.10$\pm$0.05) $\times 10^{42}$}\\
		$t_R$ (days) & 17 & 10.5 & \multicolumn{2}{c}{11.6$\pm$1.3}\\
		$M_{\rm ej}$ (M$_{\odot}$)& 10$\pm$1 & 1.8$\pm$0.8 & 8.3$\pm$2.6 &4.3$\pm$1.3 \\
		$E_{\rm K}$ (ergs) & (50$\pm$10) $\times 10^{51}$ & (2$\pm$1) $\times 10^{51}$ & (33.4$\pm$14.4) $\times 10^{51}$ & (6.1$\pm$3.9) $\times 10^{51} $  \\
		$M(^{56}\textrm{Ni})$ (M$_{\odot}$) & 0.43$\pm$0.05 & 0.2$\pm$0.04 & 0.34$\pm$0.09 & 0.42$\pm$0.08  \\
		\hline 
		\multicolumn{5}{c}{Bolometric}\\
		\hline
		$\tau$ (days) & 21.7$\pm$0.5 & 16.6$\pm$0.5 & \multicolumn{2}{c}{16.8$\pm$1}\\
		$v$ (km\,s$^{-1}$)& 20,000$\pm$1000 &16,000$\pm$1000 &\multicolumn{2}{c}{17,000$\pm$1500}\\
		$L_{\rm bol}$ (ergs\,s$^{-1}$) & 8.32 $\times 10^{42}$ & 5.58$\times 10^{42} $& \multicolumn{2}{c}{(7.7$\pm$1.4) $\times 10^{42}$}\\
		$t_{\rm bol}$ (days)& 15 & 9.6 & \multicolumn{2}{c}{13.4$\pm$2.3}\\
		$M_{\rm ej}$ (M$_{\odot}$) &10$\pm$1 & 1.8$\pm$0.8 & 5.1$\pm$0.9 & 2.0$\pm$0.3\\
		$E_{\rm K}$ (ergs) & (50$\pm$10) $\times 10^{51} $ & (2$\pm$1) $\times 10^{51}$ & (18.5$\pm$6.6) $\times 10^{51}$ & (2.5$\pm$1.4) $\times 10^{51} $ \\		
		$M(^{56}\textrm{Ni})$ (M$_{\odot}$) & 0.43$\pm$0.05 & 0.2$\pm$0.04 & 0.36$\pm$0.1 & 0.36$\pm$0.08\\
		\hline
	\end{tabular}
\end{table*}

We propagate errors in $\Delta m_{15}(R)$, the values of the analogue $M_{\rm ej}$ and $E_{\rm K}$, and on measuring the velocities through the equations to obtain an uncertainty for each parameter estimate.  We note that the largest contribution to the error budget comes from the errors in the quantities of the analogues, not from anything measured from the PTF10qts light curve.  There is a large discrepancy between the values of both quantities when using the two different analogues.

The amount of nickel produced can be estimated from the peak bolometric luminosity following the assumptions of \citet{Arnett:1982cv}. Assuming a constant bolometric correction from the $R$ band as in \citet{Drout:2011iw}, we can instead use the K-corrected magnitude in $R$. All three SNe have different rise times, so we introduce a correction to account for the varying number of $e$-folding times for primarily $^{56}$Ni, which has a half-life of 6.08\,days, and also for the decay product $^{56}$Co ($t_{1/2} = 77.23$\,days), assuming that there is no $^{56}$Co produced in the SN explosion itself. For an $R$-band luminosity $L_R$ and a nickel mass $M(^{56}\textrm{Ni})$, we find the relation
\begin{eqnarray}	
	L_R \propto M(^{56}\mathrm{Ni}) \Bigg (\frac{E_{\mathrm{\footnotesize{Ni}}}}     {\tau_{\mathrm{\footnotesize{Ni}}}}e^{-\frac{t_R}{\tau_{\mathrm{\footnotesize{Ni}}}}} + \frac{E_{\mathrm{\footnotesize{Co}}}}{\tau_{\mathrm{\footnotesize{Co}}}} \frac{\tau_{\mathrm{\footnotesize{Co}}}}{\tau_{\mathrm{\footnotesize{Co}}}-\tau_{\mathrm{\footnotesize{Ni}}}} \left[e^{-\frac{t_R}{\tau_{\mathrm{\footnotesize{Co}}}}} - e^{-\frac{t_R}{\tau_{\mathrm{\footnotesize{Ni}}}}} \right] \Bigg), 
\end{eqnarray}

\noindent where $t_R$ is the rise time in the $R$ band, and $\tau_{\mathrm{Ni}}$ and $\tau_{\mathrm{Co}}$ are the respective mean lifetimes for $^{56}$Ni and $^{56}$Co, where $\tau_i = t_{1/2} / \ln 2$. $E_{\mathrm{\footnotesize{Ni}}}$ and $E_{\mathrm{\footnotesize{Co}}}$ are the energies released by a unit mass of Ni and Co, respectively. The energy per decay is 1.7\,MeV for $^{56}$Ni and 3.67\,MeV for $^{56}$Co. As for the parameters estimated in Section \ref{sec:rband}, we use the values measured for SN\,1998bw and SN\,2006aj to provide two estimates of the nickel mass which we can then combine. The individual values for each SN are given in Table \ref{tab:phys-params}.  

This estimate is significantly different from the value obtained using the relationship in \citet{Drout:2011iw} ($\sim 0.2$\,M$_{\odot}$), despite the fact that PTF10qts is not unusual in either its $\Delta m_{15}(R)$ or $M_R$ values (see their Figure 22). This is because their relation relies purely on the absolute magnitude of the supernova at maximum and does not take into account differences in rise times.  In this study we see that PTF10qts has a similar peak magnitude in $R$-band to SN1998bw, but the rise time is $\approx 5.5$\,days shorter.  This would imply a much reduced nickel production in PTF10qts which is not reflected in the \citet{Drout:2011iw} estimation.

The fact that we obtain an even lower value with the \citet{Drout:2011iw} formula is curious; however it also predicts a lower nickel mass for SN1998bw at 0.34$M_{\odot}$.  The value for SN2006aj is in good agreement with that obtained from the modelling - 0.19$M_{\odot}$.  We attribute this to the fact that SN1998bw has a much longer rise time than all of the supernovae used in \citet{Drout:2011iw}, whereas SN2006aj has a more typical rise time.

The simplification of assuming $L \propto M(^{56}\mathrm{Ni}) / \tau$ is not appropriate for comparing supernovae with significantly different rise time or where the supernovae deviate from parabolic light curves where $\tau \propto 1/\Delta m_{15}(R)$.  We can see this is not the case for both $R$-band and bolometric light curves in Figures \ref{fig:lc_all} and \ref{fig:bolLC}.

These results clearly show it is not possible to use just the $R$-band to determine the physical parameters of this supernova and so we would caution the extension of the \citet{Drout:2011iw} relations to other supernovae, in particular where the rise time is poorly constrained or differing from $\approx$10 -- 12 days.  Instead we now focus on the generation of a bolometric light curve.

\subsection{Bolometric Light Curve}\label{sec:bolLC}

We  combine photometric and spectroscopic data to construct a pseudo-bolometric\footnote{We use the term pseudo-bolometric as the light curve we generate is
from the UV to NIR only and cannot be described as truly bolometric as
it excludes contributions at wavelengths outside this region,
particularly gamma-rays.} light curve, as this will remove the assumption that the bolometric corrections from the $R$ band for PTF10qts and either analogue are the same at all phases. 

Bolometric fluxes were computed from the six spectra by integrating their dereddened signal in the interval 4000--8500\,\AA.  As when calculating the K-corrections, we extend the 25 August (TNG) and 9 September (KPNO) spectra in the red to cover this range of wavelengths. We have also computed bolometric fluxes from the photometry at all epochs in which at least three bands were covered.  After correcting for Milky Way reddening, we converted them to fluxes according to \citet{Fukugita:1996jr}, and then splined and integrated them in the observed 4000--8500\,\AA\ range.  In the rest frame, the red boundary of this integration interval corresponds to $\sim 7800$\,\AA; thus, we have increased all bolometric fluxes by 15\% to account for the ultraviolet (UV) and near-infrared contributions (based on comparisons to other SNe that have been observed accurately both in the optical and near-infrared).  Considering the uncertainty related to this assumption and the  lack of UV information, we associate an uncertainty of 20\% with each bolometric flux.  

\begin{figure}
	\begin{center}
		\includegraphics[width=\columnwidth]{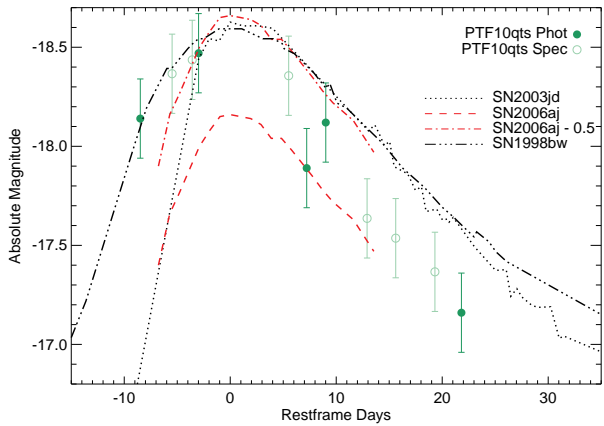} 
		\caption{The bolometric light curve of PTF10qts calculated from spectroscopic points or from photometry.  Also shown for comparison are light curves of SN\,1998bw, SN\,2006aj, SN\,2006aj $- 0.5$\,mag, and SN\,2003jd.} 
		\label{fig:bolLC} 
	\end{center}
\end{figure}

When combining the datapoints generated by these two different routes, we noted that the spectroscopically-derived points were systematically offset by a small amount to brighter magnitudes than the photometrically-derived points.  We attribute this offset to inconsistencies in the two methods used to derive the individual points.  To align the spectroscopically-generated points, we used the bolometric light curve of SN\,1998bw (itself generated via the photometric route) and fitted it to just the photometrically-derived data points of PTF10qts, allowing a temporal ``stretch'' and constant magnitude shift up and down.  Treating this warped light curve as a template, we then used $\chi^2$ minimisation to apply a constant shift to the spectroscopically-derived data to bring them in line with the photometrically-dervived points.  The final bolometric light curve is reported in Figure \ref{fig:bolLC}, where phases are plotted relative to the date of bolometric maximum, which occurs 1.84 rest-frame days after $R$-band maximum.

Comparing the shapes of bolometric lightcurves is another approximate way to examine the physical similarity of PTF10qts to other SNe Ic-BL: supernovae with similar physical properties will have similarly-shaped lightcurves.  In Figure \ref{fig:bolLC}, we show PTF10qts with the bolometric lightcurves of other SNe Ic-BL so that the dates of maximum align.  We see that the bolometric light curve of PTF10qts is most similar to that of SN\,1998bw, although the later points of PTF10qts may decline slightly faster, implying a lower nickel mass in PTF10qts.  SN\,2006aj is also a good match around maximum if it is made brighter by 0.5\,mag, although the light curve is narrower, so we would expect a higher kinetic energy and nickel mass in PTF10qts than SN\,2006aj.  We can use these observations as a sanity check when deriving physical properties from the bolometric light curve.  We also show that SN\,2003jd, which is a good match at some spectroscopic phases, is a poor match to the bolometric light curve before maximum brightness, again showing that spectroscopic similarity does not always mean the physics of the supernova explosion are the same.

We estimate the physical parameters using the relationships discussed in Section \ref{sec:rband}, but now using the bolometric quantities.   With bolometric data significantly before maximum brightness, we can switch to using $\tau$, the light-curve width, instead of just the post-maximum $\Delta m_{15} \propto 1/\tau$, which we have shown to be only an approximation.  We define $\tau$ to be the width at peak magnitude minus 0.5\,mag.  This should better reflect the differences between the SN light curves because, as Figure \ref{fig:bolLC} shows, after maximum the slopes of SN\,1998bw, SN\,2006aj, and PTF10qts are very similar, but before maximum, they differ significantly.  For PTF10qts, SN\,1998bw, and SN\,2006aj, we measure $\tau = (16.8\pm1,~21.7\pm0.5,~16.6\pm0.5)$\,days, where now PTF10qts is much less similar to SN\,1998bw and more like SN\,2006aj.

We measure the quantities when using both the SN\,1998bw and SN\,2006aj bolometric light curves, and these results are given in Table \ref{tab:phys-params}.  We again see how important it is to choose an analogue which matches both the spectroscopy and the light curve, as the estimates of the physical parameters based on SN\,1998bw and SN\,2006aj do not agree.  This is due to the different values of $E_{\rm K}/M_{\rm ej}$ and mass of $^{56}$Ni for the two analogues.  We take the weighted mean of the two analogues as the best estimate of the physics of PTF10qts: $M_{\rm ej} = 2.3 \pm 0.3$\,M$_{\odot}$ and $E_{\rm K} = (3.2 \pm 1.4)\times 10^{51}$\,ergs.  We also derive a nickel mass of $M(^{56}\textrm{Ni}) = 0.36 \pm 0.07$\,M$_{\odot}$.  The measurements of the ejecta mass and the kinetic energy are lower than using just the $R$ band, and the nickel mass is slightly higher.  We note that these estimates are similar to those for SN\,2010ah \citet{Mazzali:2013jn}, but the spectra are very different.

To estimate the zero-age main sequence (ZAMS) mass of the progenitor, we use the models of \cite{Sugimoto:1980hz} and assume a remnant mass of 2\,M$_{\odot}$ as in their models. PTF10qts corresponds to a progenitor star with a ZAMS mass of $\sim 20\pm2$\,M$_{\odot}$.

\subsection{Nebular Spectrum}

\begin{figure}
	\begin{center}
		\includegraphics[width=\columnwidth]{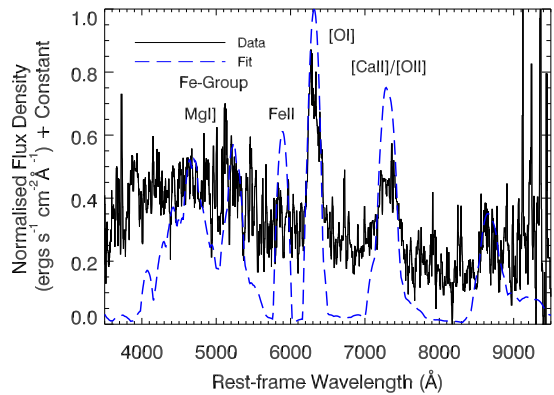} \caption{A spectrum of PTF10qts obtained at the Keck 10\,m telescope 230 days after $R$-band maximum.  A continuum has been subtracted from the data to account for host-galaxy contamination.  The dashed line is a fit to the spectrum.} \label{fig:neb-spec} 
	\end{center}
\end{figure}

We can also estimate the nickel mass from the nebular spectrum, which was obtained with LRIS at the Keck-I telescope 230 rest-frame days after $R$-band maximum. This is shown in Figure \ref{fig:neb-spec} with a continuum subtracted from it. Also shown is a synthetic spectrum.  The observed spectrum has low signal-to-noise ratio, so the resultant model fit parameters should not be used to draw any firm conclusions.  We used a code for the synthesis of nebular spectra as described by \citet{Mazzali:2001gz}.  The synthetic spectrum was obtained using $M(^{56}\textrm{Ni}) = 0.35\pm0.1$\,M$_{\odot}$, which is in good agreement with the estimate from the bolometric light curve. The red part of the spectrum also appears to indicate a low oxygen mass ($\sim0.7$\,M$_{\odot}$) in the SN, which would support the lack of detection in the post-maximum spectra (Figure \ref{fig:spec-14}); however, the blueshifted profile of the [\ion{O}{i}] emission suggests that the line may not yet be optically thin. The oxygen mass may therefore be underestimated, although we tried to take this into account in the model by requiring a stronger line than the observed one.

\subsection{Comparison to Other SNe~Ic}

Table \ref{snicbl} contains a compilation of all SNe~Ic-BL published in the literature for which physical parameters have been derived, as well as a few intermediary cases in the region between normal SNe~Ic and SNe~Ic-BL. Those SNe with which GRB events have been associated are marked by an asterisk. For SN\,2010bh we have used the models of \cite{Sugimoto:1980hz} to infer the progenitor properties from the published energetics.  PTF10qts is unremarkable among this type of SN in terms of the kinetic energy and ejecta mass, but the nickel mass is toward the higher end of the observed range.

\begin{figure}
	\begin{center}
		\includegraphics[width=\columnwidth]{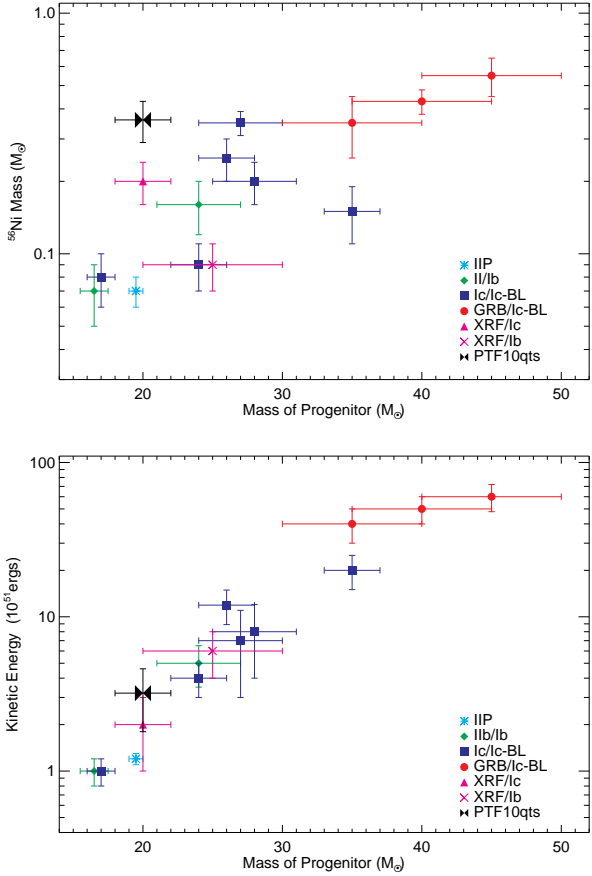} \caption{A reproduction of the plots from \citet{Mazzali:2013jn} with the addition of new data. The objects in the plot are (in order of ascending progenitor mass) \emph{SN~IIP} -- SN\,1987A; \emph{SNe~IIb/Ib} -- SN\,1993J, SN\,2003bg; \emph{SNe~Ic/Ic-BL} -- SN\,1994I, SN\,2002ap, SN\,2010ah, SN\,2003jd, SN\,2004aw, SN\,1997ef; \emph{GRB/SNe~Ic-BL} -- SN\,2003dh, SN\,1998bw, SN\,2003lw; \emph{XRF/SN~Ic} -- SN\,2006aj; and \emph{XRF/SN~Ib} -- SN\,2008D. PTF10qts is shown as the bowtie symbol.} \label{fig:phys-params} 
	\end{center}
\end{figure}

To explore this more fully, in Figure \ref{fig:phys-params} we compare PTF10qts to the trends published by \citet{Mazzali:2013jn} for energetic SNe and hypernovae where all physical parameters and the progenitor mass have been determined.  There appears to be a strong relation between the progenitor mass and the kinetic energy of the SN, and PTF10qts lies on this trend.  For example, SN\,2006aj has the same progenitor mass, and a similar kinetic energy is derived from the bolometric light curve.  The relationship between the mass of synthesised $^{56}$Ni is much looser, and PTF10qts lies away from the apparent trend, producing more nickel than would be expected for its progenitor mass.  In fact, PTF10qts has an ejected nickel mass comparable to those events classified as hypernovae.  We thus call PTF10qts a nickel-rich Type Ic-BL SN.  

	 \label{sub:search_for_GRB}

Although PTF10qts is not spectroscopically similar to SN\,1998bw, it is still photometrically similar and the event was clearly energetic. We used Interplanetary Network (IPN) data to search for a possible GRB companion to PTF10qts in case $\gamma$-rays had been detected by any of the orbiting satellites. The IPN includes Mars Odyssey, Konus-Wind, RHESSI, INTEGRAL (SPI-ACS), Swift-BAT, Suzaku, AGILE, MESSENGER, and Fermi (GBM).

The date of the PTF10qts explosion is uncertain; we know only the first detection of the SN, 5 August 2010. The observed rise time of PTF10qts is estimated to be $12.7\pm1.5$\,days.  We searched for a GRB around 16\,days before PTF10qts maximum light (allowing for any delay between a GRB and the emergence of the SN). This corresponds to a date range of 1--5 August 2010.

During this period, six bursts were detected by the nine spacecraft of the IPN. During the same period there were also 14 unconfirmed bursts which have been excluded from further analysis. The sample also excludes bursts from known sources such as anomalous X-ray pulsars and soft gamma repeaters.

Of these six bursts, three were observed with the coded fields of view of the Swift-BAT or INTEGRAL IBIS instruments, which have a positional accuracy of several arcminutes. These bursts were inconsistent with the position of PTF10qts. Two were observed either by the Fermi GBM alone, or by the Fermi GBM and one or more near-Earth spacecraft. The GBM error contours are not circles, although they are characterised as such, and they have at least several degrees of systematic uncertainties associated with them. Since no other confidence contours are specified, it is difficult to judge accurately the probability that any particular GBM burst is associated with the SN. In this analysis, we have simply multiplied the $1\sigma$ statistical-only error radius by 3 to obtain a rough idea of the $3\sigma$ error contours. One further event was observed by Konus and MESSENGER, and in this case the probability that this burst was due to PTF10qts is 0.04, excluding this as burst as conincident with the supernova.

The total area of the localisations of the six bursts was $\sim 0.04 \times 4\pi$\,steradians. This implies that there is a very low probability of finding an unassociated gamma-ray source coincident with our SN during the time window we are investigating. 

There is another approach to the probability calculation. Since only 0 or 1 GRBs in our sample can be physically associated with the SN, we can calculate two other probabilities. The first is the probability that, in our ensemble of six bursts, none is associated by chance with the SN. Let $P_i$ be the fraction of the sky which is occupied by the localisation of the $i^{th}$ burst. Then the probability that no GRB is associated with the SN is
\begin{eqnarray}
	\label{eq:nogrb} P(\mathrm{No\:GRB}) &=& \prod_i (1 - P_i). 
\end{eqnarray}

\noindent For our sample, this probability is 0.96. 

The second probability is that any one burst is associated by chance with the SN, and that all the others are not:
\begin{eqnarray}
	\label{eq:onegrb} P(\mathrm{One\:GRB\:by\:chance}) &=& \sum_i P_i \prod_{i \not= j} (1-P_j). 
\end{eqnarray}

\noindent For our sample, this probability is 0.004.

This analysis covers a very narrow range of dates for any potential GRB burst. If we extend the search period to the 30 days preceding the first optical detection of PTF10qts, there is still no statistically significant detection of any $\gamma$-rays associated with the SN event. In light of this, we assume that we have not detected any $\gamma$-rays associated with PTF10qts.

\section{Conclusions}\label{sec:conc}

We have presented optical follow-up data for the Type Ic-BL supernova PTF10qts, discovered at $z=0.0907$ by the Palomar Transient Factory. We find that the $R$-band light curve of PTF10qts is not a good representation of the bolometric light curve; hence, we used photometric and spectroscopic data to produce a pseudo-bolometric light curve from which to estimate the physical parameters of the SN explosion.  

PTF10qts appears to be a SN~Ic-BL from a progenitor of $\sim20M_{\odot}$, which is a smaller mass than some other SN~Ic-BL events, such as SN\,1998bw, SN\,2003dh and SN\,2003lw for which the progenitors are all believed to be $>35M_{\odot}$. However, PTF10qts produces a similar amount of $^{56}$Ni to these events, which are all associated with GRBs. A search of IPN data found no evidence for gamma-rays associated with the supernova event though.  PTF10qts falls on the general trends of SNe~Ic in terms of the relation between progenitor mass and kinetic energy, but for its ZAMS mass of $\sim20$\,M$_{\odot}$, it produced more $^{56}$Ni than would be expected. This is evidenced by its luminous light curve, but its narrower lightcurve width when compared to SN\,1998bw ($\tau = 21.7$\,days compared to $\tau = 16.8$\,days).  We note that the $^{56}$Ni masses we obtained by analogy with SN\,1998bw and SN\,2006aj using the $R$-band light curve (line 7 of Table \ref{tab:phys-params}) are different from those calculated via the bolometric light curve (line 14 of Table \ref{tab:phys-params}).  This indicates that the $R$-band light curve is not a completely reliable proxy for the bolometric light curve, and the latter is preferable when evaluating physical parameters.

We would caution the use of physical relationships based on monochromatic light curves for use as anything other than a first approximation.  This is because assumptions such as constant opacity, constant bolometric correction and $L \propto 1/\tau$ are oversimplifications.  In this study we have compared two methods using $R$-band and bolometric data.  We find that the bolometric methods is more suitable, but is still only an approximation.  A constraint on the time of explosion is required for this to provide anything other than a lower limit on the nickel mass.  The physical parameters of a supernova explosion of this type can \emph{only} be determined with full modelling of the light curve and spectra.

We encourage future observations of similar objects discovered early and with light curves and spectral coverage across the entire UV-optical-infrared range in order to better understand their nature.

\section*{Acknowledgements}

We acknowledge financial contributions from contract ASI I/016/07/0 (COFIS), ASI I/088/06/0, and PRIN INAF 2009 and 2011. PTF is a collaboration of Caltech, LCOGT, the Weizmann Institute, LBNL, Oxford, Columbia, IPAC, and UC Berkeley. Collaborative work between A.G. and P.A.M. is supported by a Minerva grant. The Weizmann PTF membership is supported by the ISF via grants to
A.G. Joint work of A.G. and S.R.K. is supported by a BSF grant. A.G. also acknowledges support by grants from the GIF, EU/FP7 via ERC grant 307260, ``The Quantum Universe'' I-Core program by the Israeli Committee for planning and budgeting, the Kimmel award and the Lord Sieff of Brimpton Fund.  A.V.F.’s group at UC Berkeley has received generous financial assistance from Gary and Cynthia Bengier, the Christopher R. Redlich Fund, the Richard and Rhoda Goldman Fund, the TABASGO Foundation, and NSF grant AST-1211916.   JMS is supported by an NSF Astronomy and Astrophysics Postdoctoral Fellowship under award AST-1302771.  E.O.O. is incumbent of the Arye Dissentshik career development chair and is grateful to support by a grant from the Israeli Ministry of Science and the I-CORE Program of the Planning and Budgeting Committee and The Israel Science Foundation (grant No 1829/12).

We thank the very helpful staffs of the various observatories (Palomar, Lick, KNPO, TNG, Keck) at which data were obtained. The W. M. Keck Observatory is operated as a scientific partnership among the California Institute of Technology, the University of California, and NASA; it was made possible by the generous financial support of the W. M. Keck Foundation. M. T. Kandrashoff and J. Rex assisted with the Lick observations. We are grateful to the following contributors to the IPN for support and sharing their data: I.~G.~Mitrofanov, D.~Golovin, M.~L.~Litvak, A.~B.~Sanin , C.~Fellows, K.~Harshman, and R.~Starr (for the Odyssey team), R.~Aptekar, E.~Mazets, V.~Pal'shin, D.~Frederiks, and D.~Svinkin (for the Konus-Wind team), A.~von Kienlin and A.~Rau (for the INTEGRAL team), T.~Takahashi, M.~Ohno, Y.~Hanabata, Y.~Fukazawa,  M.~Tashiro, Y.~Terada,  T.~Murakami, and K.~Makishima (for the Suzaku team), T.~Cline, J.~Cummings, N.~Gehrels, H.~Krimm, and D.~Palmer (for the Swift team), and V.~Connaughton, M.~S.~Briggs, and C.~Meegan (for the Fermi GBM team). K.H.~acknowledges NASA support for the IPN under the following grants: NNX10AI23G (Swift), NNX09AV61G (Suzaku), NNX09AU03G (Fermi), and NNX09AR28G (INTEGRAL).

\nocite{Iwamoto:2000dq} \nocite{Nakamura:2001gw}
\nocite{Mazzali:2002bf} \nocite{Mazzali:2003df} \nocite{Valenti:2008dz} \nocite{Mazzali:2006dk} \nocite{Taubenberger:2006cc} \nocite{Drout:2011iw} \nocite{Mazzali:2006bm} \nocite{Pian:2006ho} \nocite{Young:2010gv} \nocite{Sahu:2009is} \nocite{Pignata:2011hc} \nocite{Berger:2011ee} \nocite{Corsi:2011fq}  \nocite{Sanders:2012dc} \nocite{Bufano:2012ke} \nocite{Cano:2011jl}  \nocite{OlivaresE:2012jf} \nocite{Deng:2005aa} \nocite{Milisavljevic:2013aa}

\bibliographystyle{mn2e} 
\bibliography{bib_171012} \clearpage \thispagestyle{plate} \plate{Opposite p.~8} 
\begin{table*}
	\vbox to220mm{
	\vfil Landscape SN~Ic-BL table to go here. \caption{} 
	\vfil} \label{snicbl} 
\end{table*}

\end{document}